\documentclass[aps,notitlepage,prl,twocolumn,superscriptaddress,a4paper,floatfix,showpacs]{revtex4-1}
\usepackage{epsfig}
\usepackage{amsmath,amssymb,amsfonts}
\usepackage{hyperref}
\usepackage{mathrsfs}
\usepackage{bbm}
\usepackage{slashed}
\usepackage{graphicx}
\usepackage{verbatim}
\usepackage{url}
\usepackage[normalem]{ulem}
\usepackage{mathptmx}

\usepackage[usenames]{color}

\newcommand{\be}{\begin{equation}}
\newcommand{\ee}{\end{equation}}

\newcommand{\bi}{\begin{itemize}}
\newcommand{\ei}{\end{itemize}}
\newcommand{\bea}{\begin{eqnarray}}
\newcommand{\eea}{\end{eqnarray}}
\newcommand{\ud}{\mathrm{d}}

\graphicspath{{figures/}}

\definecolor{darkgreen}{rgb}{0.2,0.7,0}
\definecolor{amethyst}{rgb}{0.6, 0.4, 0.8}

\usepackage[T1]{fontenc} \usepackage[latin1]{inputenc}

\begin{document}
\title{Quantum quenching of radiation losses in short laser pulses}

\author{C.~N.~Harvey}
\email[]{cnharvey@physics.org}
\affiliation{Department of Physics, Chalmers University of Technology, SE-41296 Gothenburg, Sweden}
\author{A.~Gonoskov}
\email[]{arkady.gonoskov@chalmers.se}
\affiliation{Department of Physics, Chalmers University of Technology, SE-41296 Gothenburg, Sweden}
\affiliation{Institute of Applied Physics, Russian Academy of Sciences, Nizhny Novgorod 603950, Russia}
\affiliation{Lobachevsky State University of Nizhni Novgorod, Nizhny Novgorod 603950, Russia}
\author{A.~Ilderton}
\email[]{anton.ilderton@plymouth.ac.uk}
\affiliation{Department of Physics, Chalmers University of Technology, SE-41296 Gothenburg, Sweden}
\affiliation{Centre for Mathematical Sciences, Plymouth University, PL4 8AA, UK}
\author{M.~Marklund}
\email[]{mattias.marklund@chalmers.se}
\affiliation{Department of Physics, Chalmers University of Technology, SE-41296 Gothenburg, Sweden}

\begin{abstract}
Accelerated charges radiate, and therefore must lose energy. The impact of this energy loss on particle motion, called radiation reaction, becomes significant in intense-laser matter interactions, where it can reduce collision energies, hinder particle acceleration schemes, and is seemingly unavoidable. Here we show that this common belief breaks down in short laser pulses, and that energy losses and radiation reaction can be controlled and effectively switched off by appropriate tuning of the pulse length. This "quenching" of emission is impossible in classical physics, but becomes possible in QED due to the discrete nature of quantum emissions.
\end{abstract}

\maketitle

Continual advances in achievable laser power has spurred renewed interest in using intense light to study fundamental predictions of classical and quantum electrodynamics (QED)~\cite{Marklund:Review,DiPiazza:Review,Marklund:photon,Heinzl:biref,King:Nature}.
One cornerstone of such experiments is the collision of laser beams with particle bunches~\cite{Chen,E144}. Particle motion in intense fields is inherently non-linear, in particular due to radiation reaction (RR) which is the impact of energy loss on particle motion. RR can reduce collision energies~\cite{Fedotov:classical}, hinder particle acceleration schemes~\cite{RDR1,Malka,DiPiazza:Review}, and is seemingly unavoidable. Much work has gone into demonstrating that RR, long thought negligible, must now be accounted for in order to accurately model state-of-the-art high intensity laser-matter interactions~\cite{DiPiazza:Review,Burton:2014wsa,Gonoskov:2015}. In contrast, we will show here that we can control, and effectively turn off RR by tuning the laser pulse length. We will also present a realisable experimental setup, requiring only modest parameters,  with which to observe the effect and so demonstrate a possibility to control quantum processes in intense light-matter interactions.

Consider then the collision of an electron beam with an intense laser. The trajectory and energy evolution of the electrons follows, in QED, a probability distribution which is typically centred on lower energy losses than is predicted by classical physics. The reason is that while accelerated electrons must radiate continuously according to classical mechanics, the stochastic nature of quantum processes allows electrons to penetrate into the pulse before losing any significant energy to emission~\cite{Shen:1972,Duclous,Blackburn:straggling}. The purely quantum effect we present here is that electrons can interact with the entire laser pulse, but pass through it without losing energy to hard photons.  This is forbidden in classical physics, where radiative losses and recoil effects are continuous phenomena~\cite{DiPiazza:Review,Burton:2014wsa}, but is made possible by tuning the laser pulse length and exploiting the discrete nature of quantum emission. Due to the latter, there is always a nonzero chance for the electrons to not emit any photons \textit{of sufficient energy to significantly back-react on the electron}. To roughly estimate when this phenomena may be significant, we consider the elastic scattering probability $P_0$ that the electron does not emit radiation, $P_0 = \exp(- P_1)$, where $P_1$ is the probability of one emission~\cite{IR}. ($P_1$ is infrared finite in laser backgrounds~\cite{RitusReview} and soft emission does not cause any significant back-reaction on the electrons, see Appendix for details). $P_0$ is exponentially damped with both intensity and pulse duration, so signatures of quantum effects, though present, are normally obscured in e.g.~long laser pulses~\cite{Harvey:2016}; once the electrons emit they quickly lose energy, entering a classical regime. Thus in order to identify parameters for observing our effect we first consider short pulses.

\begin{figure*}[t!!!]
\includegraphics[width=1\textwidth]{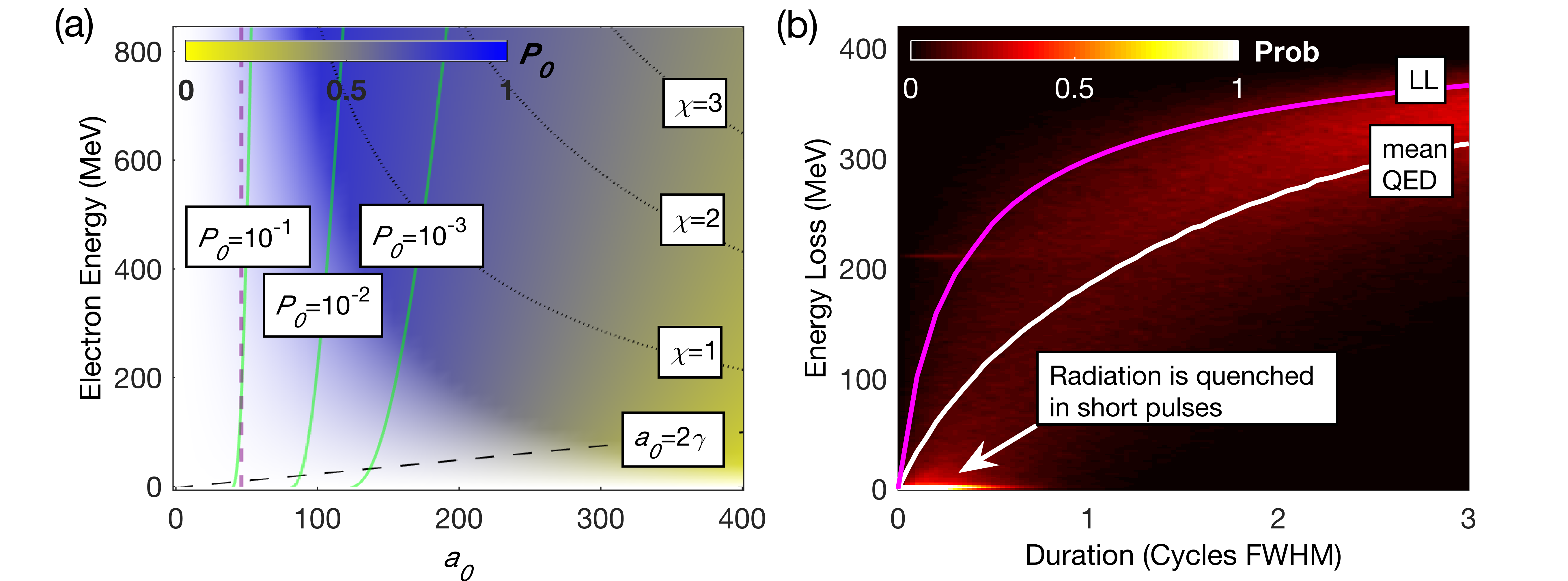}
\caption{\label{fig:Chris_test_plot}
{\bf a.)}  The no-emission probability $P_0$ (colour) in a one-cycle pulse as a function of initial electron energy and peak laser electric field ($E$) in relativistic units $a_0 = e E \lambda/2\pi m c^2$ (for electron charge $e$, mass $m$ and $c$ is the speed of light). The laser has wavelength $\lambda=0.82\mu m$ and spot size $w_0=5\mu m$. As the probability of not emitting is unity if, say, $a_0=0$ (no laser), we restrict attention to the radiation dominated regime~\cite{RDR1,Piazza:exact} where the parameter $\epsilon_\text{rad} = (2/3)\alpha \omega a_0^2\gamma/m\geqslant1$ and significant recoil effects are expected: the colour is therefore faded out for $\epsilon_\text{rad}<1$. The optimal parameter region (blue) is then $a_0\in \{100\ldots 200\}$ and electron energies in the 100s of MeV.  This region overlaps with that where $\chi<1$: we work throughout in this regime, so that pair production need not be considered~\cite{RitusReview}. Also shown is the line $a_0=2\gamma$ above which classical RR causes reflection of the electron~\cite{DiPiazzaRR:2009}. {\bf b.)} Probability density of the electron energy loss in QED for initial electron energy 420MeV and $a_0=200$. The corresponding peak intensity is $1.1\times 10^{23}$ W/cm$^2$, slightly beyond the state-of-the-art~\cite{Yanovsky:2008} but within reach of upcoming facilities~\cite{Vulcan10,ELI}. The average energy loss in QED (white line) grows more slowly than the classical prediction (pink line) of the Landau-Lifshitz equation (`LL', see Appendix). The bright yellow-white region in the lower left-hand corner shows that there is a high probability for the elections to pass through a short pulse without losing energy to emission. This is `quenching'.
}
\end{figure*}

In Fig.~\ref{fig:Chris_test_plot}a we plot $P_0$ in a pulse of FWHM duration 2.7~fs, corresponding to one optical cycle~\cite{Lin:2006}, as a function of laser intensity and initial electron energy. The implied optimal parameters for intensity and energy are then verified in Fig.~\ref{fig:Chris_test_plot}b where the electron energy loss is plotted as a function of pulse length. We simulate the laser-particle interactions using well-tested Monte-Carlo routines~\cite{Harvey:2015,Green:2015,Gonoskov:2015}, see Appendix.  As expected, the mean energy loss grows much more slowly in QED than classical physics predicts but, more significantly, the bright region in the lower left-hand corner of Fig.~\ref{fig:Chris_test_plot}b shows that there is a high probability for the elections to pass through a short pulse without losing energy to emission.

\begin{figure}
\includegraphics[width=0.98\columnwidth]{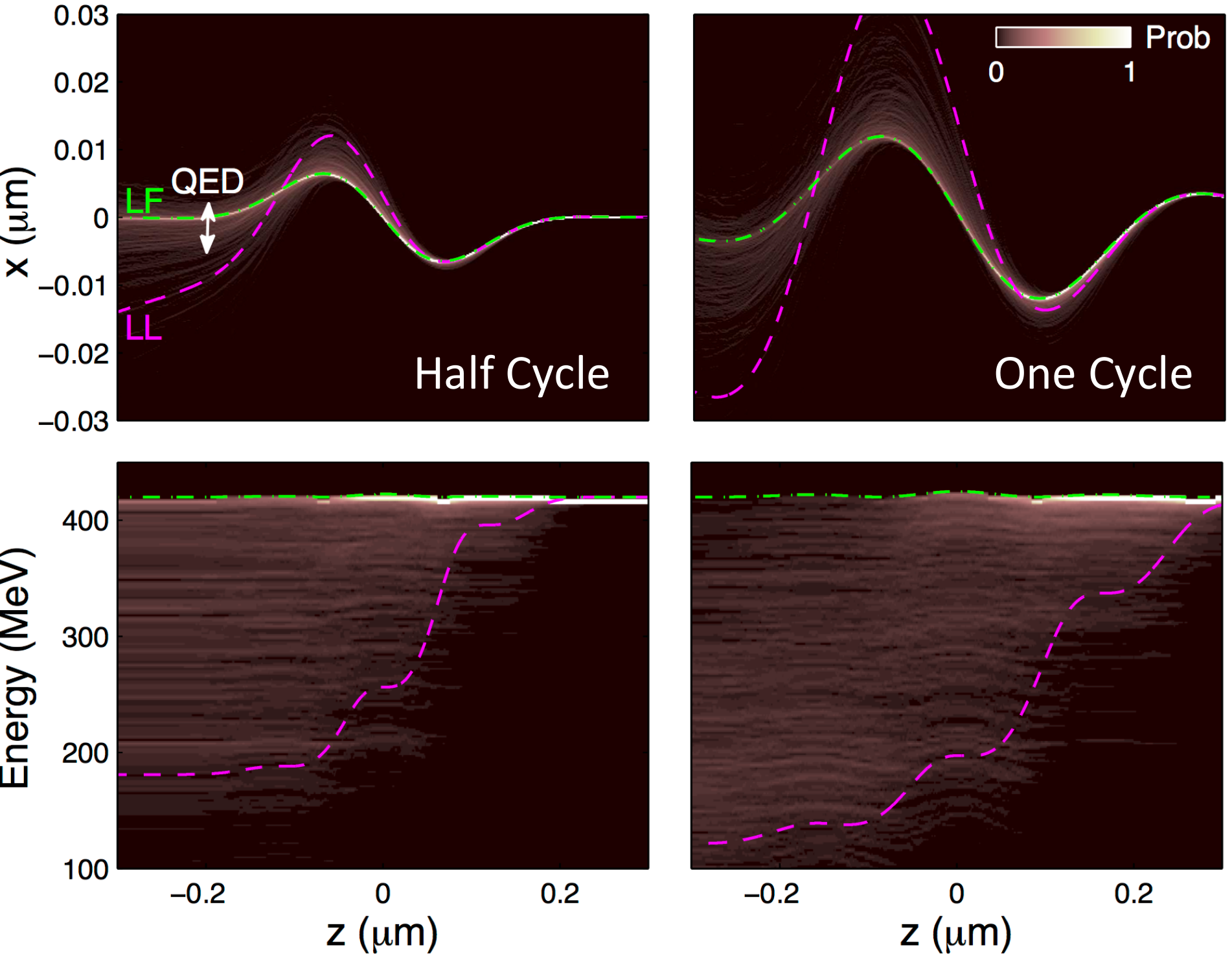}
\caption{Trajectories and energies of a single electron, incident from the right and passing through the centre of an ultra-short, focussed laser pulse of duration 0.5 cycles (1.4~fs, left) and one cycle (2.7~fs, right). Other parameters as in Fig.~\ref{fig:Chris_test_plot}b. The coloured region shows the QED probability distribution, calculated from 1000 simulations with the same initial conditions. (For clarity any region containing~$>200$ electrons is coloured as if it contained 200.) This distribution does not follow classical predictions which include RR, shown as a purple dashed line, but instead is visibly centred (white) on the Lorentz force curve shown with a dashed green line: this trajectory is by definition absent of RR. The lower panels show that, since the pulse is short, there is a high probability for the electrons not to emit until they are past the pulse peak. Even for the few electrons which subsequently emit, it is too late for their motion to be significantly affected. \label{fig:density} }
\end{figure}

Figs.~\ref{fig:density} show the dynamics of a single relativistic electron passing through the centre of an intense pulse. The upper panels show that the electron probability density is not centred on the radiating trajectory expected classically, but on the Lorentz force trajectory, i.e.~a no-emission, no-recoil path: even though the electron clearly interacts with, and is accelerated by, the pulse, RR is effectively switched off.

We can make an analogy between this effect and that of chemical fluorescence quenching~\cite{Lakowicz}, in which excited electrons can move between molecules without emission of photons. This radiation-free transfer decreases the fluorescent intensity of a sample. Quenching mechanisms are typically short-range, with e.g.~Dexter transfer being purely quantum mechanical and exponentially damped with distance~\cite{Dexter}. Due to these similarities we refer to our effect as \emph{quenching} of radiation. Quenching is related to ``straggling''~\cite{Shen:1972,Blackburn:straggling}, that is the possibility for electrons to reach the focus of a laser pulse and emit higher energy photons than would be possible classically. Straggling is necessary but not sufficient for quenching: the latter requires the combination of {\it the effects of} straggling with a short duration pulse.

As a check we use known analytic results to recover some properties of Fig.~\ref{fig:Chris_test_plot}. The invariant $\chi = e \hbar \sqrt{p.F^2.p}/m^3 c^4$ (where $F_{\mu\nu}$ is the electromagnetic field tensor and $p^\nu$ the electron momentum), parameterises the importance of strong field quantum effects. These are present even at $\chi\ll 1$~\cite{Nerush20117}, and in this regime the emission probability is well approximated by ${\ud P_1}/{\ud t} \simeq {1.44\alpha} \chi/{m\gamma} $ for $\alpha$ the fine structure constant $\gamma$ the electron gamma factor~\cite{RitusReview}. Given the high particle energy considered, it suffices to integrate over the plane-wave profile through the centre of the pulse, which gives $P_1 \simeq {1.44\alpha} a_0 \times 4.69$, where $a_0$ is the peak field strength and the final numerical factor comes from integrating over the pulse profile (see Appendix). This implies that the curve $\exp(-P_1)=1/10$ should be independent of $\gamma$, i.e.~approximately vertical, at $a_0\simeq 47$, in excellent agreement with the plotted results, see the purple dashed line in Fig.~\ref{fig:Chris_test_plot}a.
 
\begin{figure}[t!]
\includegraphics[width=1\columnwidth]{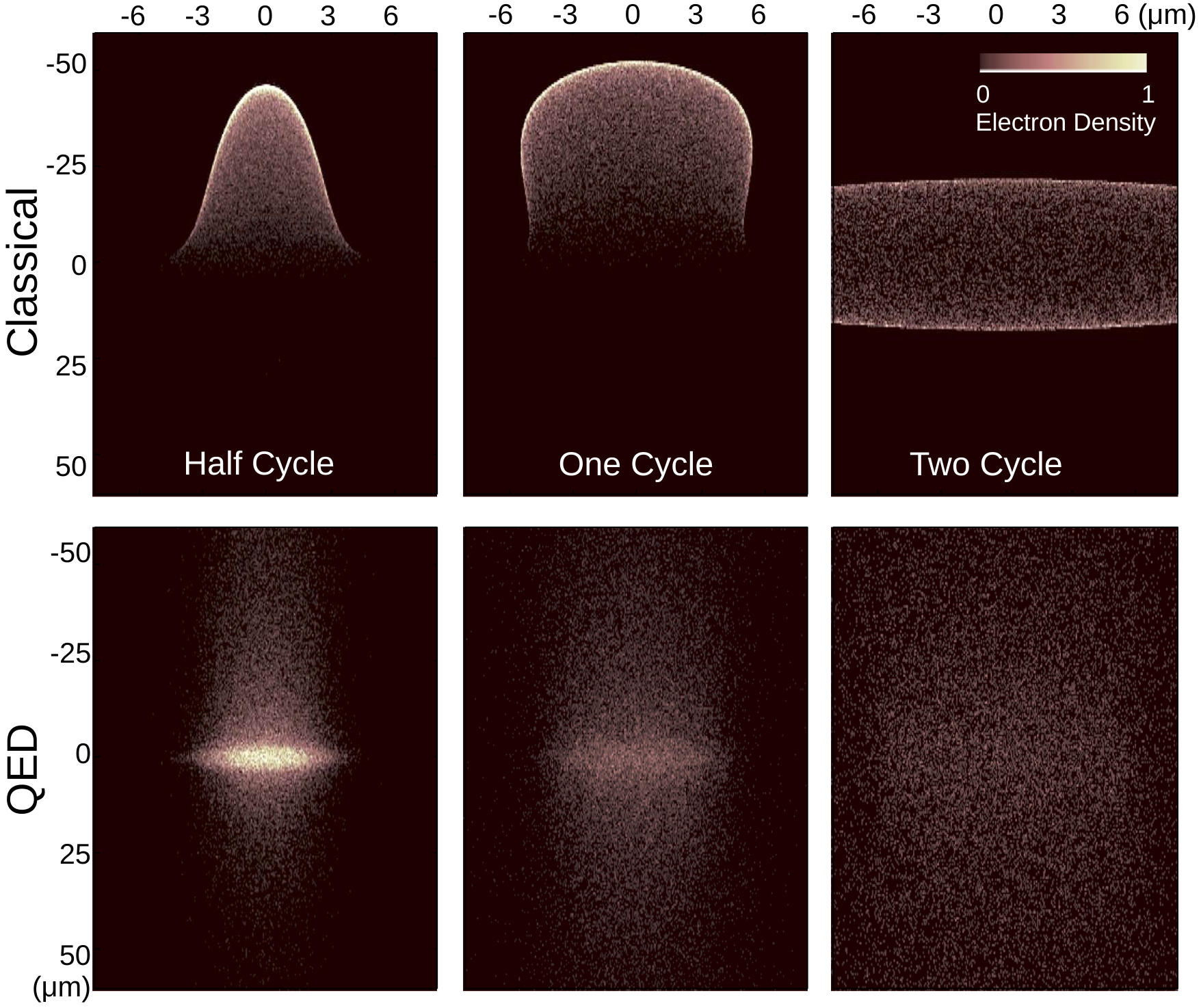}
\caption{Quenching in the collision of an electron bunch with laser pulses of different durations: the plots show the final electron distribution on a lanex screen $1mm$ behind the focus. Classical predictions (top) show an asymmetric spatial distribution for short pulses. In the quantum results (bottom) the bright spot is formed by electrons which have traversed the pulse without significant emission or energy loss. The effect becomes less prominent as pulse length increases. For two cycles the classical distribution becomes symmetric, as one would expect, while in the quantum case stochastic transverse spreading~\cite{Green:2014} causes the bunch to spread out to a wider spot size than the classical bunch. The electron bunch initially had Gaussian distributions in energy, $420 \pm 0.35$ MeV, and spacial position, $2\mu m$ FWHM. Laser intensity $a_0=200$, focal spot size $w_0=5\mu m$.  (Parameters are similar to those achievable at ELI-NP~\cite{ELI}.)\label{fig:screens}}
\end{figure}

With this confirmation, we can increase complexity by considering collisions between the laser and, now, a realistic electron bunch. Fig.~\ref{fig:screens} shows the final distribution of electrons on a screen positioned 1~mm directly behind the laser. Parameters are chosen to optimise quenching: increasing $a_0$ widens the deflection angle of the classical electrons, while increasing $\gamma$ makes the final spot sizes smaller and more collimated.  Classical predictions suggest that electrons radiate a substantial amount of their energy in the front tail of the pulse, causing them to slow down and be deflected.  For longer durations, both the classical and quantum spectra are symmetric, though the latter exhibit typical stochastic spreading effects. For pulse durations of one cycle and below, the classical deflection is asymmetric. However, a quantum calculation shows that the electrons are now not deflected, but hit the centre of the screen: quenching allows electrons to enter, and cross, field regions which are forbidden according to classical physics. This provides a clear signal which can be pursued experimentally: we look for electrons in places where there is zero classical background. (This same principle can also enhance signals of vacuum birefringence~\cite{Holger:VacBiref}.) Note that, in contrast to straggling, we do not need to detect the emitted photons.
 
\begin{figure}
\includegraphics[width=1\columnwidth]{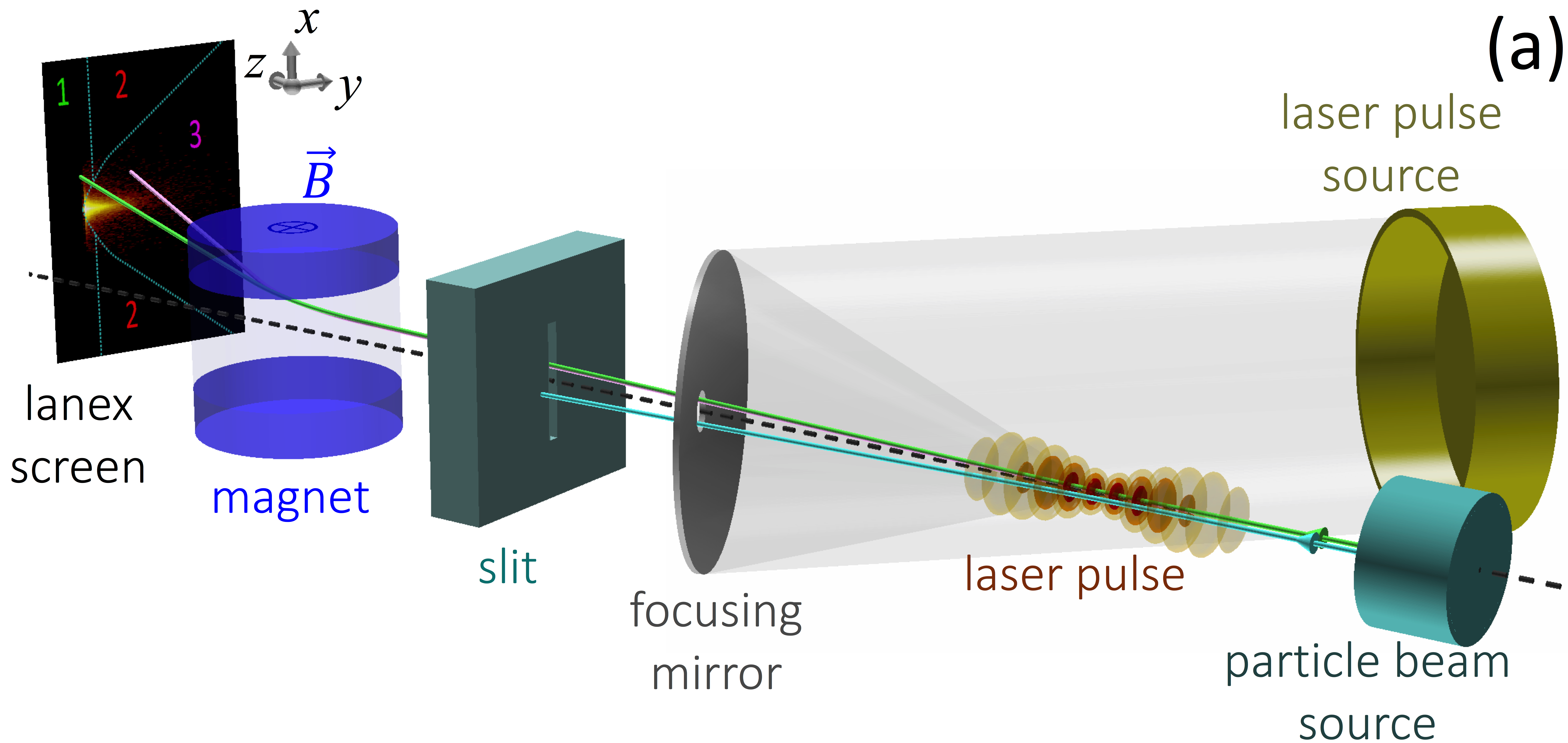}
\includegraphics[width=1\columnwidth]{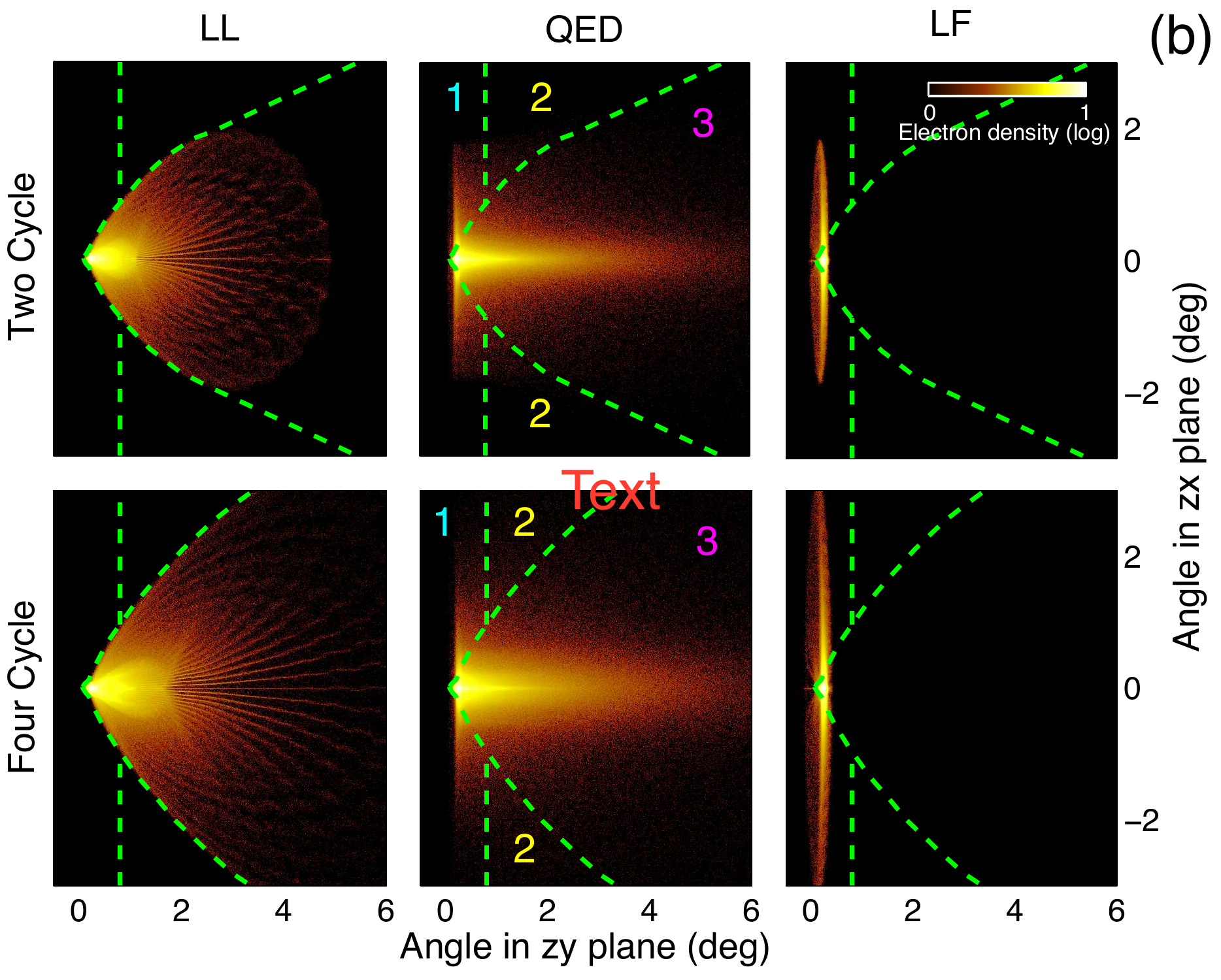}
\caption{
\label{fig:expt} a.) The proposed experimental setup for observing quenching. b.) Lanex screens for 2 and 4 cycle pulses. Peak intensity $a_0=50$ (50TW total power), wavelength $820\,$nm, tightly focussed using an f/1 optic. A $0.32\,$T, $15\,$cm magnet is placed $1\,$m behind the slit, which deflects electrons in the $y$-direction according to their energy. The lanex screen is placed 50~cm behind the magnet and angled at $45^\circ$. The electron beam energy is 100 MeV (with 0.1\% spread) and the beam is large enough so that precise synchronisation would not be a concern in an experiment. The quantum distribution is very different from that predicted classically (`LL'), where electrons are confined to region 3. Quenched electrons, which have been deflected by the laser but not lost energy, will be delivered to region~1. Electrons can also appear in the non-classical region~2, but due to stochastic spreading~\cite{Green:2014}: they have lost energy during the interaction. The boundary of region 1 is composed of the boundaries of the classical region 3 and the no-recoil Lorentz force prediction (`LF').}
\end{figure}

Although the short pulses above are currently out of experimental reach, the stochastic nature of quantum emission means that quenching is still present even in longer pulses, although it is harder to observe because the electrons undergo classical-like cooling~\cite{Pom,Neitz:2013qba,Yoffe:2015mba} before they emerge from the pulse. We therefore present a simple experimental setup, illustrated in Fig.~\ref{fig:expt}a, which filters out noise and mimics the short-pulse regime. (Note that field focussing will play an essential role here, in contrast to the parameters used for Fig.~\ref{fig:Chris_test_plot}.) Here an electron bunch is brought into collision with a, now, tightly focussed laser pulse. The electrons subsequently pass through a slit, then through a magnet, before being dumped onto a lanex screen. The setup is designed such that any electrons for which radiation is quenched will have properties distinct from all others and will populate a particular portion of the screen, free from noise. The laser is linearly polarised in the $x$--direction.  The collision occurs along the $z$--axis. The magnet is orientated such that electrons are fanned out in the $y$--direction according to their energy, before hitting the screen. The main bulk of the electrons that miss the centre of the laser focus will form a bright spot on the screen. Electrons passing close to the most intense part of the pulse can, for tight focussing, receive a significant deflection in the $x$-direction. (This is larger than e.g.~ponderomotive pushing could provide for electrons in the pulse periphery.) Electrons which have lost energy to emission will be fanned out by the magnet in the $y$-direction. However, electrons which have been significantly deflected in the $x$--direction but which have not been significantly deflected in $y$ have necessarily passed through the high-field region but, due to quenching, did not lose energy. The screen area they occupy cannot be populated by stochastic spreading, as all such electrons will have lost energy and will therefore be deflected in the $y$-direction. A possible source of noise would be electrons which have gained momentum in the $y$-direction due to either tight focussing effects, ponderomotive pushing, or stochastic spreading. These electrons would give a false reading, but are removed by the slit after exiting the laser, so never reach the magnet or screen.

\begin{figure}
\includegraphics[width=1\columnwidth]{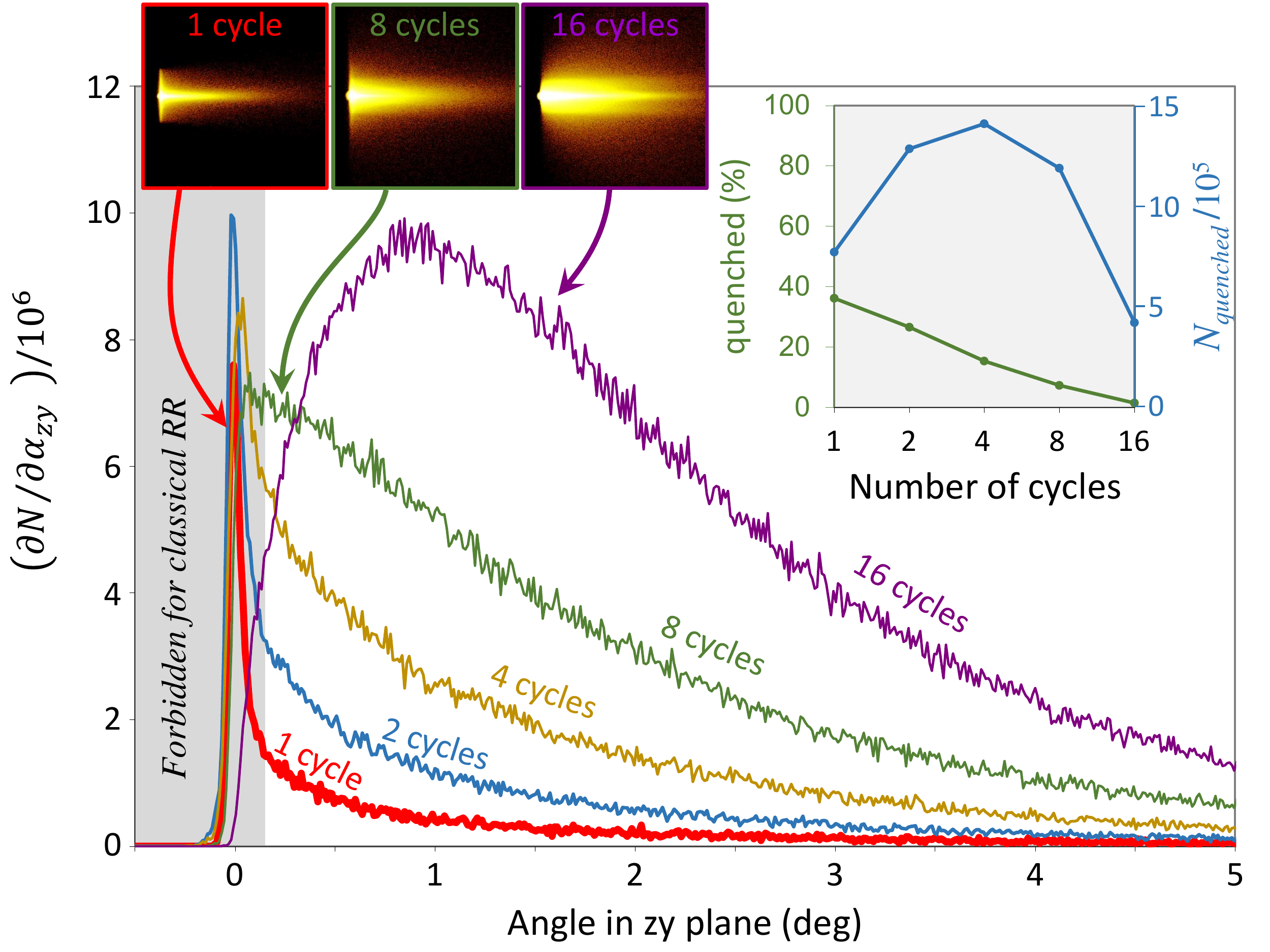}
\caption{\label{fig:5}
3D QED PIC simulation results showing the density of electrons deflected in the $x$-direction by $>0.5$ deg as a function of the $zy$ angle for pulse durations 1, 2, 4, 8 and 16 cycles (FWHM). The light grey zone (< 0.15 deg) demarks the classically forbidden region. The upper right inset shows the dependency of the total number of quenched electrons (from the grey region) on the number of cycles (right hand scale, blue), and as a percent of all particles deflected by more than 0.5 deg in the $x$-direction (left hand scale, green).}
\end{figure}

Simulated experimental results are shown in Fig.~\ref{fig:expt}b for parameters within reach of current facilities~\cite{ELI}. (The vector beam model used above has a ring singularity~\cite{Fedotov,Harvey:focusing}; this was irrelevant for our previous parameters, but could cause problems for the wide electron beam used here. Therefore we propagate the pulse using the quantum particle-in-cell code ELMIS3D~\cite{Gonoskov:2015}.) The results demonstrate the feasibility of the proposal.  The quantum prediction is vastly different to that of classical RR, signalling the presence of quantum effects, and the electron distribution more closely mimics the Lorentz force (no recoil) prediction. The electrons which have been deflected by the laser but not lost significant energy, demonstrating quenching, are deposited in region 1 of the screen. These electrons occupy an area of the screen which is free from classical noise. The electrons which have undergone quantum stochastic spreading hit the screen in region 2; they also occupy a non-classical, but different, area of the screen.  (The filamentation in the classical LL distribution is a consequence of the tight pulse focussing.) Thus quenching can be observed even in longer laser pulses, and at currently available parameters. To underline this, Fig.~\ref{fig:5} shows the proportion of electrons experiencing quenching as a function of pulse duration. Even for 16 cycles the number of quenched electrons is non-negligible ($\sim1.5\%$). The detection of quenching is therefore within reach of existing facilities. The inset provides another experimental signature: a peak in the absolute number of electrons quenched as a function of pulse duration.

In conclusion, we have shown that quantum effects allow for an electron to be accelerated and decelerated by a short laser pulse without emitting hard photons; in effect radiation reaction is turned off, and electrons can follow Lorentz force trajectories, barely perturbed by emission and energy losses. One of the goals of new intense-laser facilities, such as ELI-NP, is to observe such fundamental quantum phenomena~\cite{ELI}. It is anticipated that high intensities and short pulse durations will come hand-in-hand at future facilities~\cite{Mourou:2011}, and indeed quenching is most prominent in short pulses. It can though also be observed in longer pulses using currently available parameters. Finally, we remark that determining the properties of high-intensity pulses remains an open and challenging problem: we show in the Appendix that quenching has a potential application here, as it can be used to measure carrier envelope phase.

\begin{acknowledgments}
\textit{The authors are supported by the Knut~\&~Alice Wallenberg Foundation, the Swedish Research Council, grants 2012-5644 and 2013-4248 (CH~and MM), the Olle Engkvist Foundation, grant 2014/744 (AI), the EU's Horizon 2020 research and innovation programme under the Marie Sk\l odowska-Curie grant No.~701676. (AI), the Russian Science Foundation, project no.~16-12-10486 (AG), and the Russian Foundation for Basic Research project no.~15-37-21015 (AG).  AI~thanks J.J.~Phillips for useful discussions. Part of the simulations were performed on resources provided by the Swedish National Infrastructure for Computing (SNIC) and HPC2N.}
\end{acknowledgments}

\appendix
\section{Appendix A: Simulation details}
\subsection{Vector beam}

Simulations using the vector beam model were conducted with the code SIMLA~\cite{Green:2015}, based on the same assumptions and probabilistic procedures as now routinely employed in large-scale QED-PIC simulations~\cite{Elkina:2010up,Ridgers:2014,Gonoskov:2015,Vranic2016}. SIMLA is a single particle code, i.e.~Coulomb field effects in electron bunches are neglected. This is sufficient for our purposes as we do not consider dense bunches, and the dominant field is always that of the intense laser.

Paraxial models of focussed pulses are not valid for subcycle durations.We begin by using instead a vector beam model~\cite{Lin:2006} in which no unphysical DC field components appear for subcycle durations. While the field profile is similar to that of a focussed Gaussian beam, the price to pay is that there is a ring singularity in the field at finite waist size~\cite{Fedotov}. This has no impact on the majority of our calculations as our electron beams pass well within the ring, but for the tightly focussed beam required for Fig.~4 of the text we must do better, see below, and Fig.~\ref{FIG:VECTORSUPP}.

Adopting units such that $\hbar=c=1$, we define the peak laser field amplitude in terms of the dimensionless parameter $a_0$. At high intensity, $a_0\gg 1$ (intensity $\sim a_0^2$, with $a_0=1$ corresponding to $\sim10^{18}$~W/cm$^2$ at optical frequency) the formation length of photon emission is of order $\lambda/a_0\ll\lambda$, where $\lambda$ is the laser wavelength~\cite{RitusReview}. As a result, processes may be calculated in a locally constant approximation, and multi-vertex process (e.g.~two photon emission) become `factorisable' into sequential single vertex events~\cite{Nikishov:1964zza,RitusReview}. The rate of photon emission, call it $\Gamma$, is then a function of position due to the spatio-temporal variation of the laser fields, and due to the electron motion. For explicit expressions see~\cite{Elkina:2010up,Gonoskov:2015}. Quantum processes become more probable as the value of the `quantum efficiency parameter', $\chi$, increases from zero -- for an electron of momentum $p$ or a photon of momentum $k'$ in a field $F_{\mu\nu}$ the quantum efficiency parameters are
\be
	\chi_e = \frac{\sqrt{e^2p.F^2.p}}{m_e^3} \;, \qquad \chi_\gamma = \frac{\sqrt{e^2k'.F^2.k'}}{m_e^3} \;,
\ee
and are again local functions. For $\chi<1$ effects such as pair production are exponentially suppressed~\cite{RitusReview}, but we can still have significant quantum effects in photon emission~\cite{RitusReview,Nerush20117,DiPiazza:Review,QRR}, which is the situation of interest here.

\begin{figure}[t!]
\includegraphics[width=\columnwidth]{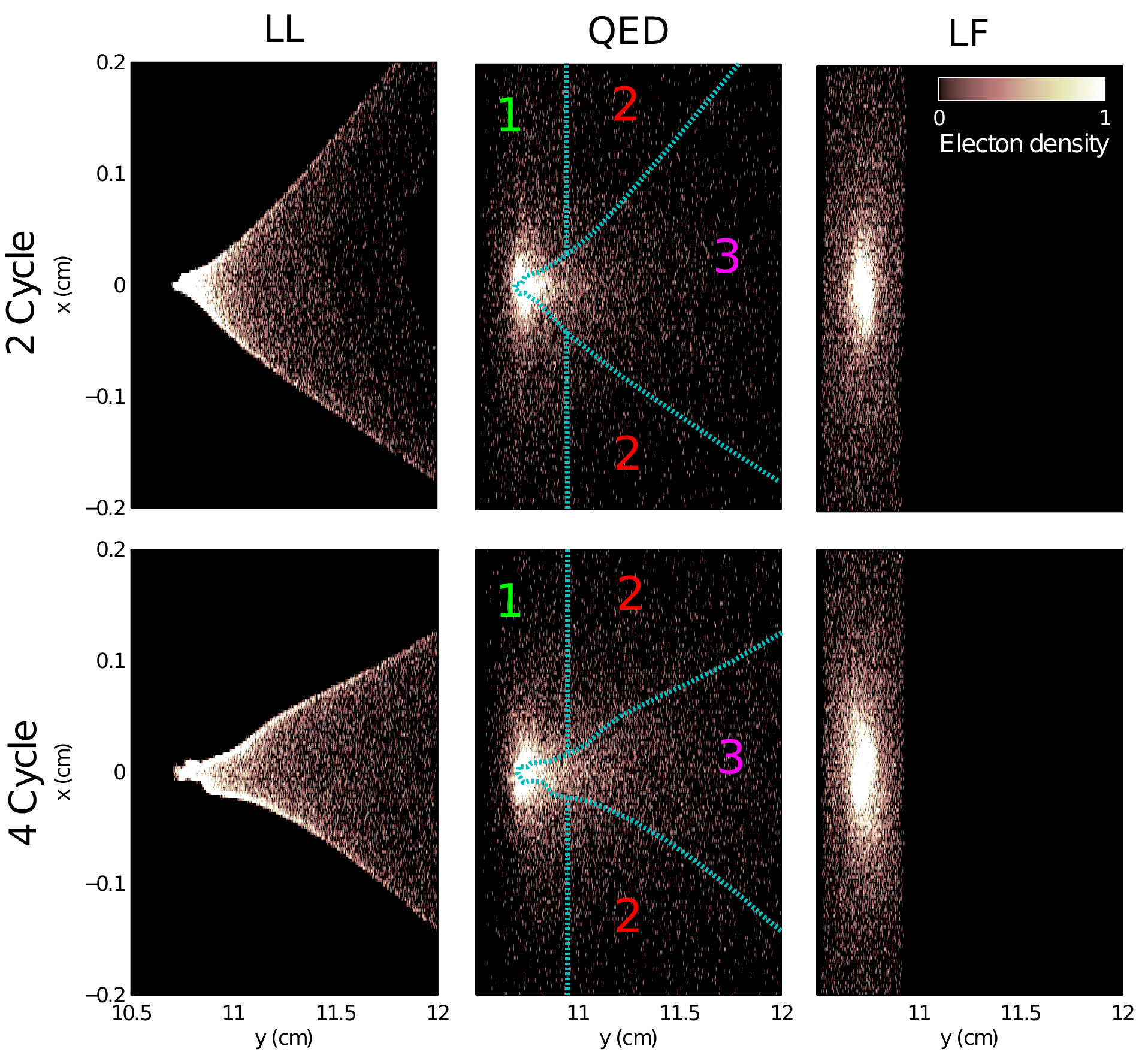}
\caption{
\label{FIG:VECTORSUPP} Observation of quenching using the vector beam model. Quenching is clearly visible as the presence of electrons in region 1, as in the main text. However, it was found that a small number of electrons in the simulation had passed close to the ring singularity of the vector beam. In order to remove any doubt in the results, the vector beam model was therefore abandoned and PIC simulations were used instead. This figure is included for completeness. Lanex screens for 2 and 4 cycle (FWHM) pulses. Peak laser intensity $a_0=140$, wavelength $\lambda=820\,$nm and waist diameter equal to $\lambda$. A $2\,$T, $15\,$cm magnet is placed $1\,$m behind the slit, which deflects electrons in the $y$-direction according to their energy. The lanex screen is placed 50~cm behind the magnet and angled at $45^\circ$. (Note that we also use $y$ to denote the deflection on the screen.) The bunch of $5\times 10^4$ electrons had length $400\,$pm, energy 600 MeV ($\pm 0.1\%$) and FWHM width equal to $2 \mu$m. Around half the electrons had a divergence angle greater than 1mrad when leaving the laser and were stopped by the slit.
}
\end{figure}

In the simulations, electrons are evolved over discrete time steps~$\Delta t$ (much shorter than the time scales of the laser field) via the Lorentz force equation. After each step the following statistical routine is used to calculate the probability of photon emission and to correct the momentum of the electron due to recoil, i.e.~to include radiation and radiation reaction in a fully quantum manner. A uniform random number $s\in[0,1]$ is generated, and emission occurs if $ s\leq \Gamma \Delta t$. Given this, a second uniform random number $\zeta\in[0,1]$ is generated and used to determine the frequency of the emitted photon (through $\chi_\gamma$) as the root of the sampling equation~\cite{Elkina:2010up,Duclous}
\be\label{stickprov}
	\zeta={\Gamma}^{-1} \int_{0}^{\chi_{\gamma}}\!\ud\chi_\gamma \frac{\ud \Gamma}{\ud\chi_\gamma} \;.
\ee
The direction of photon emission is fixed as forward relative to the emitting electron direction~\cite{Jackson, Harvey:2009}. Finally the electron is recoiled: the emitted photon momentum is subtracted from the electron momentum, imposing the conservation law $\chi_e\to \chi_e-\chi_{\gamma}$ (which applies beyond the plane wave model because at high intensity any field looks, to the particle, like a crossed field, i.e.~a constant plane wave~\cite{RitusReview}) and the simulation proceeds by propagating the photon (on a linear trajectory) and the electron (via the Lorentz equation) to the next time step, where the latter may emit again. In this way the algorithm captures all multi-photon effects \textit{at high intensity}; there are of course limitations to these methods, e.g.~that the intensity must obey $a_0\gg 1$, as we have, but these are well known and have been thoroughly investigated in the literature, see~\cite{Elkina:2010up,King:2013zw,Harvey:2015,Gonoskov:2015,QRR}.

We comment that the sampling equation (\ref{stickprov}) requires integrating over arbitrarily low emitted photon frequencies at small $\chi_\gamma$. At first glance this seems at odds with the standard QED result that photon emission probabilities are IR divergent~\cite{IR}. However, the potential singularity is weakened from $1/\omega$ to $1/\omega^{-1/3}$ for emission in a crossed field and in the LCFA model, and so becomes integrable~\cite{RitusReview}. (Codes often employ a cutoff at low $\chi$, but see~\cite{Gonoskov:2015} for improved methods.) While it is still impossible to detect arbitrarily low energy photons, soft emission is any case irrelevant here, as only hard emissions significantly affect the electron motion.

\subsection{PIC Simulation details}
For Fig.~4 of the main text we performed 3D simulations using the particle-in-cell (PIC) code ELMIS3D~\cite{gonoskov.phdthesis}. Although the Coulomb interaction between particles was negligible for our configuration, the use of a full PIC model is important to demonstrate that the expected signature is clearly detectable with realistic focusing optics. The ELMIS3D code has a spectral solver for Maxwell's equations with no numerical dispersion, in any direction, and thus perfectly reproduces field configurations with tight focusing. To simulate f/1 optics, we generated laser radiation in the far-filed zone with a spherical phase front and a uniform intensity distribution within the opening angle of $2\tan^{-1}(1/2)$. The adaptive event generator described and tested in~\cite{Gonoskov:2015} was used for the QED simulations; this approach reproduces the full spectrum of photon emission in both QED and classical regimes, without the need to impose an explicit IR cutoff.

\section{Appendix B: Analytic results and CEP}
Physics in short pulses can be sensitive to carrier envelope phase (CEP) effects. We must therefore check that CEP does not destroy or obscure quenching. We confirm here that it does not, while also showing that quenching can in principle be used to determine the CEP of a laser pulse. This investigation will also serve as a check on our numerical results.

\begin{figure}[t!!]
\includegraphics[width=0.99\columnwidth]{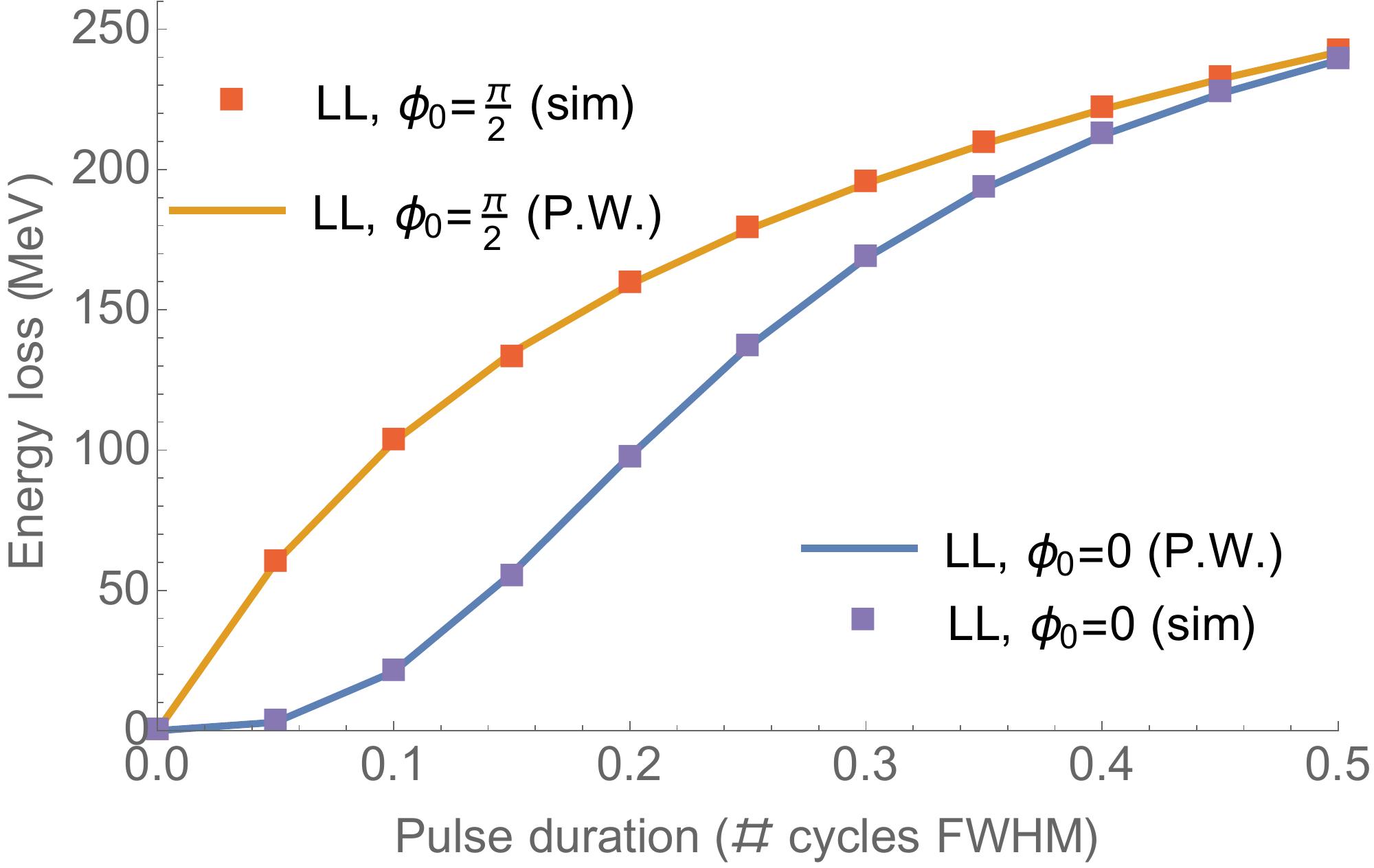} \\[10pt]
\includegraphics[width=0.99\columnwidth]{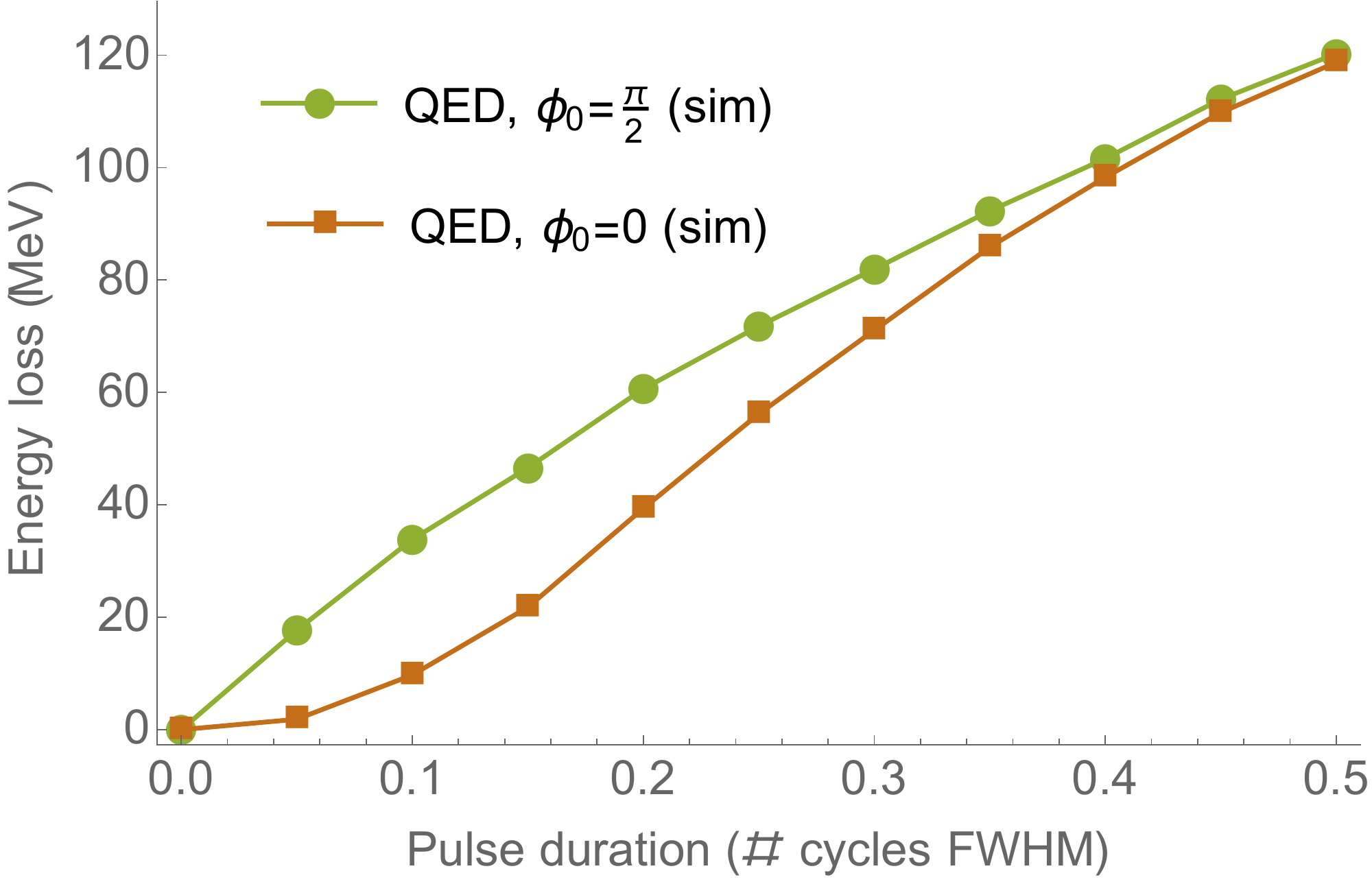}  \\[10pt]
\includegraphics[width=0.49\columnwidth]{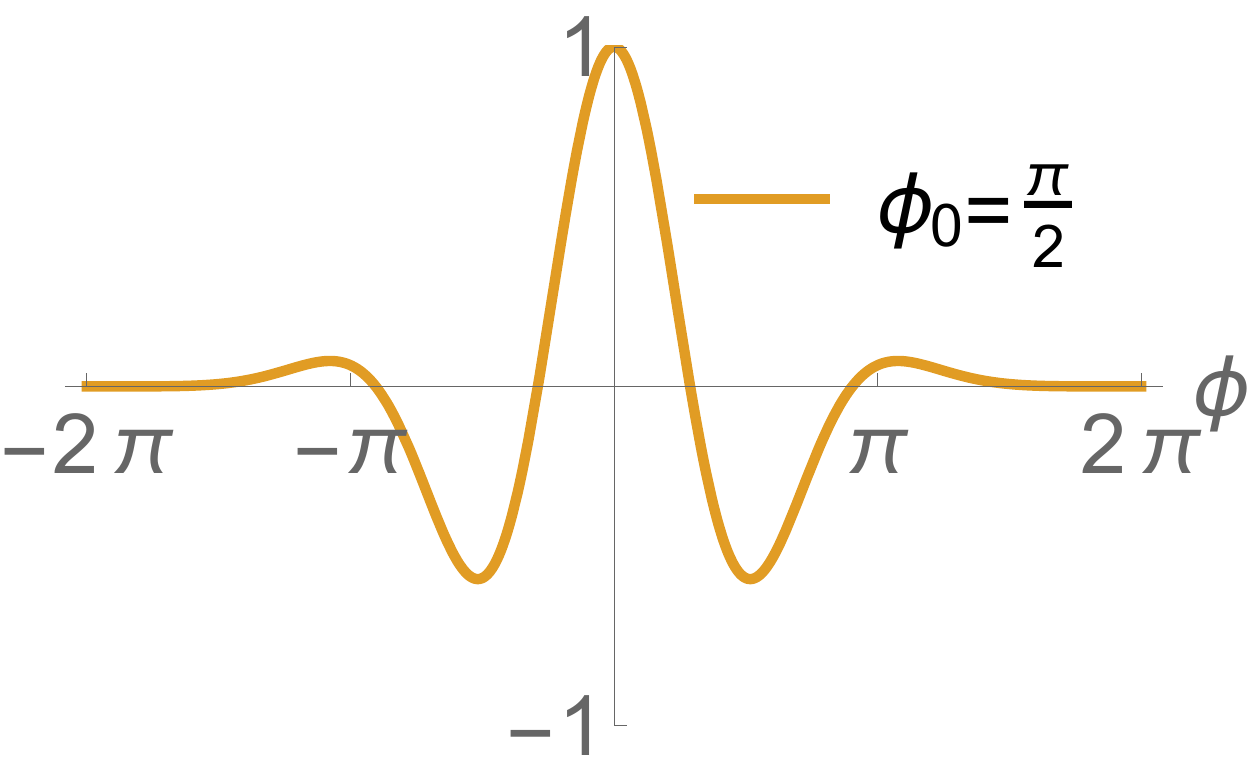}\includegraphics[width=0.49\columnwidth]{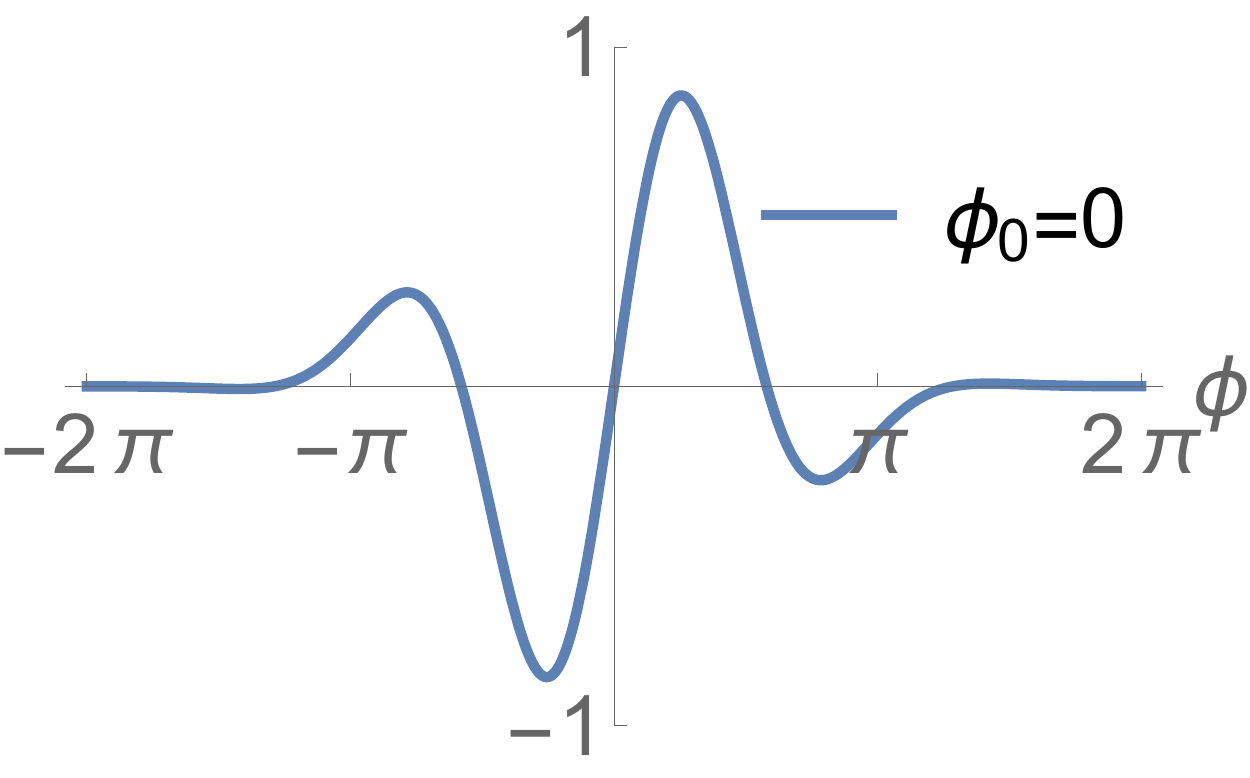}
\caption{
The impact of carrier envelope phase (CEP) on the electron energy loss, as a function of pulse duration. Parameters as in Fig.~1 of the text (initial electron energy 420MeV, intensity $a_0=200$), but with varying pulse length and two carrier phases $\phi_0=\pi/2$ and $\phi_0=0$.  \textit{Top panel:} Classical simulations (``sim'') demonstrate a sensitivity to CEP, and are confirmed by excellent agreement with the analytic solution of the LL equation using the plane wave limit of the focussed pulse (``P.W.''). This also confirms that focussing effects are negligible. \textit{Middle panel:} QED results (averaged over 10,000 simulations) show that the energy loss in the quantum theory is, for short pulses, significantly lower than the classical theory predicts, consistent with quenching. The sensitivity to CEP remains.  \textit{Bottom panels:} Example $1/2$-cycle field profiles with different carrier phases.
\label{FIG:CEP}
}
\end{figure}

For the setup behind Fig.~1 and~2 of the text, the high particle energy and collision geometry mean that beam focussing effects should be negligible, and hence the simulation results should be recoverable from a plane wave model of the laser field, which allows for analytic calculations. In this limit the electric field of the vector beam is~\cite{Lin:2006}
\be\begin{split}\label{E}
	e E(\phi) &= m a_0 \text{Im }\, A_d(\phi) A_d(0)^{-1} e^{-\phi^2/2\tau^2 + i \phi + i\phi_0} \;, \\
	A_d(\phi) &= \tau^{-2} + (1+i \phi \tau^{-2})^2\;,
\end{split}
\ee
where $\phi=\omega(t+z)$ with $\omega$ the central frequency, $\phi_0$~is the carrier phase, and $\tau$ is chosen such that the pulse contains $r$~wavelengths at FWHM, implying $\tau := \pi r /\sqrt{\log 4}$. There is no DC component and the ring singularity of the 3D beam is absent in this limit. Example pulse profiles are shown as part of Fig.~\ref{FIG:CEP}.  We note, as used in the main text, that integrating $\chi$, essentially $|E|$, over the pulse duration gives
\be
	\int\!\ud\phi\, |eE(\phi)| \simeq a_0 \times  4.69 \;.
\ee
for a single cycle pulse.

In Fig.~\ref{FIG:CEP} we compare the simulation results behind Fig.~1 and Fig.~2 of the text with analytic calculations, and simultaneously investigate CEP effects by now allowing for two different carrier phases. The analytic classical results are obtained from the exact solutions of the Lorentz force (LF) and Landau Lifshitz (LL) equations in a plane wave~\cite{Landau:1982dva,Piazza:exact}. The net energy gained by an electron traversing the profile~(\ref{E}) is zero according to the Lorentz equation. Solving the LL equation and taking the difference of the final and initial electron energies then gives the energy lost to radiation, which may be directly compared with the energy emitted by electrons in the simulation. The classical predictions match exactly with our simulation results, which also confirms that focussing effects are negligible for the considered parameters and that the ring singularity of the 3D beam is inconsequential. The impact of CEP can clearly be seen in the differences in the shape of the curves. This persists in the quantum theory: the emitted energy is clearly sensitive to CEP. However, CEP does not damage quenching: for both of the chosen carrier phases we see that quantum effects significantly reduce the energy lost to radiation in short pulses as compared to classical predictions. This is consistent with quenching in short pulses. We have also confirmed that a full QED calculation based on single photon emission in a plane wave agrees exactly with simulation results (not shown) for very short pulses. However, we have also found that multi-photon effects become significant even for half-cycle pulses at the intensities considered, and as \textit{exact} analytic results are available only for one and two-photon emission, an analytic comparison covering a wide range of pulse durations is not available. The most reasonable analytic approximation to make, which would allow multi-photon effects to be calculated semi-analytically, would amount to an LCFA approximation, which is why we instead use the well-tested~\cite{Elkina:2010up,King:2013zw,Harvey:2015,Gonoskov:2015,QRR} numerical routines described above.

%

\end{document}